\documentclass[prb,twocolumn,showpacs,preprintnumbers,amsmath,amssymb]{revtex4}
\usepackage{graphicx}% Include figure files
\usepackage{psfig}
\usepackage{dcolumn}% Align table columns on decimal point
\usepackage{bm}% bold math

\begin{document}

%\preprint{APS/123-QED}

\title{Selective dilution and magnetic properties of
 La$_{0.7}$Sr$_{0.3}$Mn$_{1-x}$\textit{M}$^\prime_x$O$_3$ (\textit{M}$^\prime$ = Al, Ti)}% Force line breaks with \\
\author {D. N. H. Nam,$^{1,2}$ L. V. Bau,$^{1,3,5}$ N. V.
Khiem,$^{4}$ N. V. Dai,$^{1}$ L. V. Hong,$^{1}$ N. X. Phuc,$^{1}$ R.
S. Newrock,$^{2}$ and P. Nordblad$^{5}$}
\affiliation{$^1$Institute
of Materials Science, VAST, 18 Hoang-Quoc-Viet, Hanoi, Vietnam \\
$^{2}$Department of Physics, University of Cincinnati, OH
45221-0011, USA\\$^3$Department of Science and Technology, Hongduc
University, Thanhhoa, Vietnam\\$^4$Department of Natural Sciences,
Hongduc University, Thanhhoa, Vietnam\\$^5$The \AA ngstr\"{o}m
Laboratory, Uppsala University, Box 534, SE 751-21 Uppsala, Sweden}
\date{\today}% It is always \today, today,
             %  but any date may be explicitly specified
\begin{abstract}
The magnetic lattice of mixed-valence Mn ions in
La$_{0.7}$Sr$_{0.3}$MnO$_3$ is selectively diluted by partial
substitution of Mn by Al or Ti. The ferromagnetic transition
temperature and the saturation moment decreases with substitution in
both series. The volume fraction of the non-ferromagnetic phases
evolves non-linearly with the substitution concentration and faster
than theoretically expected. By presenting the data in terms of
selective dilutions, the reduction of $T_\mathrm{c}$ is found to be
scaled by the relative ionic concentrations and is consistent with a
prediction derived from molecular-field theory.
\end{abstract}
% PACS, the Physics and Astronomy
                             % Classification Scheme.
%\keywords{Suggested keywords}%Use showkeys class option if keyword
                              %display desired
\maketitle
\section{\label{sec:intro}Introduction}
Perovskite manganites with the general composition ($R$,$A$)MnO$_3$
($R$: rare earth, $A$: alkali elements) are attractive to scientists
not only because of their potential applications but also due to
their very rich and intriguing physics. While most of the pristine
compounds $R$MnO$_3$ are (insulating) antiferromagnets due to
antiferromagnetic (AF) superexchange (SE) interactions between
Mn$^{3+}$ ions, the introduction of divalent alkali cations such as
Sr$^{2+}$, Ca$^{2+}$, or Ba$^{2+}$ into the composition converts an
adapted number of Mn$^{3+}$ to Mn$^{4+}$ that in turn gives rise to
the $\mathrm{Mn}^{4+}-\mathrm{O}^{2-}-\mathrm{Mn}^{3+}$
ferromagnetic (FM) double-exchange (DE) \cite{Zener} interaction.
The presence of DE couplings can turn an insulating
antiferromagnetic manganite into a ferromagnet with metallic
conductivity. It has been widely accepted that, along with lattice
distortions \cite{Millis1,Millis2, Hwang} and phase segregation
phenomena, \cite{Dagotto} the DE mechanism plays a very important
role in governing the properties of manganites.

Perovskite manganites have been intensively studied in the last
decade since the discovery of the Colossal Magneto-Resistance (CMR)
phenomenon. \cite{Helmolt} Chemical substitution has been widely
used as a conventional method to uncover the underlying physics and
to search for compositions with novel properties. Results for both
$R$- and Mn-site substitution have been quite well documented in the
literatures. For manganites with a Mn$^{3+}$/Mn$^{4+}$ ratio of 7/3,
substitution of different elements at the rare-earth site has shown
that magnetic and transport properties of many manganites depend
rather systematically on the average ionic size, $\langle
r_{A}\rangle=\sum_{i}y_{i}r_{i}$, and the ionic size mismatch
defined as $\sigma^2=\sum_{i}y_{i}r_{i}^2-\langle r_{A}\rangle^2$
(where $y_i$ is the fraction and $r_i$ the ionic radius of the $i$th
species occupying the $R$-site), \cite{Hwang, Teresa} i.e. on the
degree of GdFeO$_3$-type lattice distortion and lattice disorder.
However, despite many attempts at modifying the magnetic lattice of
Mn ions by direct substitution at the Mn-sites, the complexity
caused by too many factors that govern the properties of the
materials, no universal features have been revealed to date. In this
paper, we report that the reduction of the ferromagnetic ordering
transition temperature $T_\mathrm{c}$ of
La$_{0.7}$Sr$_{0.3}$Mn$_{1-x}$\textit{M}$^\prime_x$O$_3$ ($M^\prime=
\mathrm{Al}$, Ti) scales with the relative substitution
concentrations and is consistent with a prediction from
molecular-field theory (MFT) of the dilution of a magnetic lattice.
The substitutions with Al and Ti are \emph{selective} in nature
because Al$^{3+}$ only substitutes for Mn$^{3+}$ and Ti$^{4+}$ only
for Mn$^{4+}$. Another advantage of these substitutions is that
because Al$^{3+}$ and Ti$^{4+}$ ions do not carry a magnetic moment,
they are also expected not to participate in the magnetic
interaction.
\section{\label{sec:exper}Experiment}
In this paper,
La$_{0.7}$Sr$_{0.3}$Mn$_{1-x}$\textit{M}$^\prime_x$O$_3$ is denoted
as LSMA$_x$ for \textit{M}$^\prime$ = Al and LSMT$_x$ for
\textit{M}$^\prime$ = Ti. All the samples were prepared using a
conventional solid state reaction method. Pure ($\geq$99.99\%) raw
powders with appropriate amounts of La$_2$O$_3$, SrCO$_3$, MnO$_2$,
Al$_2$O$_3$, and TiO$_2$ are thoroughly ground, mixed, pressed into
pellets and then calcined at several processing steps with
increasing temperatures from 900 $^\mathrm{o}$C to 1200
$^\mathrm{o}$C and with intermediate grindings and pelletizations.
The products are then sintered at 1370 $^\mathrm{o}$C for 48 h in
ambient atmosphere. The final samples are obtained after a very slow
cooling process from the sintering to room temperature with an
annealing step at 700 $^\mathrm{o}$C for few hours. Room-temperature
X-ray diffraction patterns (measured by a SIEMENS-D5000 with
Cu-$K_\alpha$ radiation) show that all of the samples are
crystallized in perovskite rhombohedral structures with almost no
sign of secondary phases or remnants of the starting materials. The
crystal structures obtained for the samples are in agreement with
earlier structural studies. \cite{Qin,Hu,Kallel,Kim,Troyanchuk}
Transport and magnetotransport measurements were carried out in a
non-commercial cryostat using the standard 4-probe method. Magnetic
and magnetization measurements were performed in a Quantum Design
MPMS SQUID magnetometer and (sometimes) by a Vibrating Sample
Magnetometer (VSM).
\begin{figure}
  % Requires \usepackage{graphicx}
  \vspace{0.1in}
  \includegraphics[width=3.3in]{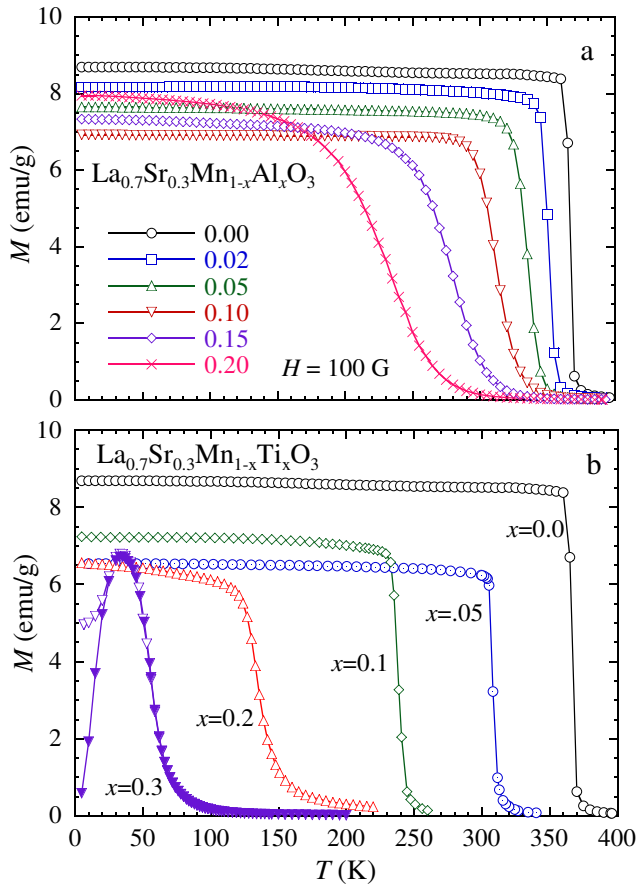}
  \caption{Field-cooled magnetization as a function of temperature for
(a)
$\mathrm{La}_{0.7}\mathrm{Sr}_{0.3}\mathrm{Mn}_{1-x}\mathrm{Al}_x\mathrm{O}_3$
and (b)
$\mathrm{La}_{0.7}\mathrm{Sr}_{0.3}\mathrm{Mn}_{1-x}\mathrm{Ti}_x\mathrm{O}_3$,
measured in $H = 100$ G. The zero-field-cooled
\textit{M}$_\mathrm{ZFC}(T)$ ($\blacktriangledown$) of the $x = 0.3$
sample is added for further discussions.}\label{Figure1}
\end{figure}
\section {Results and Discussion}
Magnetic, transport, and magnetotransport characterization of
LSMA$_x$ and LSMT$_x$ compounds has been reported previously by
other authors and can be referenced in a number of publications.
\cite{Qin,Hu,Kallel,Kim,Troyanchuk,Sawaki} Although consistent
tendencies as to the variation of e.g. the transition temperature
and saturation moment with doping concentration have been found,
there is significant scatter in the data in-between the different
studies. In the current work, the essential characteristics are
carefully measured and reexamined.
\subsection{Temperature dependent characterization}
Temperature dependent magnetization measurements, $M(T)$, in both
zero-field-cooled (ZFC) and field-cooled (FC) protocols, are carried
out for all the samples. The FC $M(T)$ curves presented in Fig. 1
clearly indicate that the substitution of Al or Ti for Mn causes the
ferromagnetic-paramagnetic (PM) transition temperature
$T_\mathrm{c}$ to drop drastically. The reduction of $T_\mathrm{c}$
in the case of Ti substitution is much more substantial. The FM-PM
transition is very sharp at small $x$ concentrations for both doping
series but becomes broader with increasing $x$. A transition is
observed for all the samples, even with LSMT$_{0.3}$ where Mn$^{4+}$
ions are supposed to be completely absent. However, as is implied by
the ZFC and FC $M(T)$ curves in Fig. \ref{Figure1}, the
LSTMT$_{0.3}$ sample is not a true ferromagnet, nor a pure spin
glass as has been suggested in Ref. 12 for this composition. This
last conclusion is also corroborated by the fact that our
frequency-dependent ac-susceptibility measurements
$\chi_\mathrm{ac}(T,\omega)$ (not shown) do not indicate a dynamic
phase transition. The $M(T)$ curves for this sample probably suggest
the existence of an AF background state with a weak ferromagnetic
component --- possible indications of a canted antiferromagnet, as
was proposed in Ref. 10. The $T_\mathrm{c}$ vs. $x$ data extracted
from the $M(T)$ curves for all samples are plotted in Fig.
\ref{Figure4} and will be discussed later in detail.

Temperature dependent transport measurements for both series show
that the resistivity, $\rho$, strongly increases with increasing $x$
while the metal-insulator (MI) transition usually observed at a
temperature $T_\mathrm{p}$ near (but lower than) $T_\mathrm{c}$ is
shifted to lower temperatures. The MI transition is observed in all
of the LSMA$_x$ samples while it can only be observed in LSMT$_x$
for $x$ up to 0.1. For LSMT$_{0.2}$, the MI transition is no longer
observed, but there still exists a slope change in the $\rho(T)$
curve at a temperature below $T_\mathrm{c}$ signalling a magnetic
contribution from the FM phase to the conductivity. In agreement
with the magnetic behavior, the LSMT$_{0.3}$ sample only shows
insulating characteristics.
\begin{figure}
  % Requires \usepackage{graphicx}
  \vspace{0.1in}
  \includegraphics[width=3.3in]{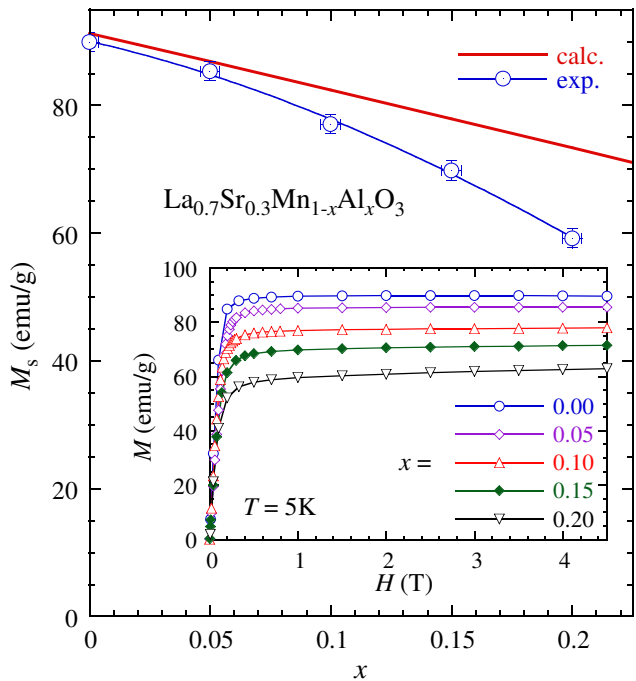}
  \caption{$\mathrm{La}_{0.7}\mathrm{Sr}_{0.3}\mathrm{Mn}_{1-x}\mathrm{Al}_x\mathrm{O}_3$:
Theoretical $M_\mathrm{s}$ vs. $x$ calculated from equation
\ref{eqn0} (bold curve without symbols) and experimental
$M_\mathrm{s}$ data (thin curve with $\odot$ symbols) extracted from
$M(H)$ measurements at 5 K (inset).}\label{Figure2}
\end{figure}
\subsection{Magnetization characterization}
\begin{figure}
  % Requires \usepackage{graphicx}
  \vspace{0.1in}
  \includegraphics[width=3.3in]{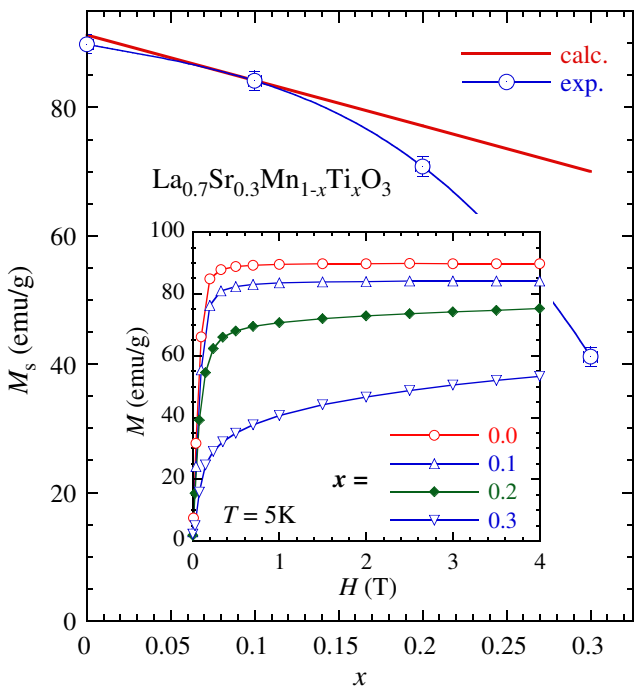}
  \caption{$\mathrm{La}_{0.7}\mathrm{Sr}_{0.3}\mathrm{Mn}_{1-x}\mathrm{Ti}_x\mathrm{O}_3$:
Theoretical $M_\mathrm{s}$ vs. $x$ calculated from equation
\ref{eqn0} (bold curve without symbols) and experimental
$M_\mathrm{s}$ data (thin curve with $\odot$ symbols) extracted from
$M(H)$ measurements at 5 K (inset).}
  \label{Figure3}
\end{figure}
The theoretical zero-temperature spin-only saturation magnetization
(in emu/g) of the LSMA$_x$ and LSMT$_x$ compounds follow
\begin{eqnarray}
M_\mathrm{s}=\left\{ \begin{array}{ll}
            (3.7-4x)10^3/(40.548-5.006x) & \mbox{if $M^\prime=\mathrm{Al}$}\\
            (3.7-3x)10^3/(40.548-1.801x) & \mbox{if
            $M^\prime=\mathrm{Ti}$}.
           \end{array}
     \right.
\label{eqn0}
\end{eqnarray}
For the LSMA$_x$ series, magnetization measurements at $T = 5$ K
(the inset of Fig. 2) indicate that the samples are substantially
saturated in an applied field of just above 1 T. At higher fields,
the LSMA$_0$ and LSMA$_{0.05}$ samples exhibit flat $M(H)$
dependencies as expected for conventional ferromagnets at high
fields and low temperatures. Closer inspection on the $M(H)$ curves
for $x\geq0.1$ shows, however, that they are quite linear in the
high-field regime up to 4.5 T with a small slope which increases
with $x$. The small slope would come from the suppression of thermal
fluctuations of the magnetization by the magnetic field. However,
the evolution of the slope of the $M(H)$ curves with $x$ in the high
field regime may signal a magnetic contribution from certain Mn ions
that do not take part in the ferromagnetic phase. In addition, for
the whole series, the measured magnetization in magnetic fields up
to 4.5 T does not reach the theoretical magnetization value and even
deviates further with increasing $x$. Based on these features, it is
presumable that segregation into FM and non-FM phases occurs in
these samples. The experimental values of the saturation
magnetization $M_\mathrm{s}$ of the FM phase presented in Fig.
\ref{Figure2} (and also Fig. \ref{Figure3} for Ti substitution) are
determined by extrapolating the linear part of the $M(H)$ curves in
the high field regime to $H = 0$. The difference between
experimental and theoretical $M_\mathrm{s}$ is then attributed to
the non-FM contributions, $M_\mathrm{non-FM}$.

Very similar results are also obtained for the LSMT$_x$ series, as
can be seen in Fig. \ref{Figure3}. Previous observations of the
decrease of $M_\mathrm{s}$ in Al and Ti substitutions in manganites
have been published and interpreted by several other authors.
\cite{Kallel,Kim,Troyanchuk,Blasco,Cao} The decrease of
$M_\mathrm{s}$ was attributed to the dilution and the weakening of
FM exchange couplings as in general. Kallel \textit{et al.}
\cite{Kallel} suggest even a change of the spin state or the orbital
ordering of the Mn ions. The deviation between the experimental
$M_\mathrm{s}$ and its theoretical values has not been adequately
considered. In a study on
La$_{0.7}$Ca$_{0.3}$Mn$_{1-x}$Al$_{x}$O$_3$ by neutron diffraction,
the effective magnetic moment per Mn ion was found far smaller than
the saturation moment and that was explained by assumptions of
magnetic inhomogeneity and structural disorder.\cite{Blasco} Our
interpretation for the $M_\mathrm{s}$ deviation in terms of magnetic
phase segregation is somehow closer to that proposed for
La$_{0.7}$Ca$_{0.3}$Mn$_{1-x}M^\prime_x$O$_3$ ($M^\prime$= Ti, Ga)
compounds by Cao \textit{et al.}, \cite{Cao} where the authors
suggest that Ti$^{4+}$ and Ga$^{3+}$ ions generate around them
non-ferromagnetic (paramagnetic or possibly antiferromagnetic)
regions.

As pointed out above, there is a growing deviation of the FM
saturation magnetization from the calculated value with increasing
substitution concentration. This observation indicates that the
substitution not only reduces the total number of Mn ions but also
raises the number of non-FM Mn ions at the expense of the FM phase.
Since the Mn$^{3+}$/Mn$^{4+}$ ionic concentration ratio is driven
away from the optimal value of 7/3 by the substitution, a certain
amount of the Mn ions may become \emph{redundant} with respect to
the DE couplings. I.e., the selective substitution on one Mn ionic
species produces the redundancy on the other Mn species. Those
redundant ions contribute to the non-FM phase. With small values of
$x$, the redundant ions are thus mostly Mn$^{4+}$ for $M^\prime=
\mathrm{Al}$ while they are Mn$^{3+}$ for the Ti substitution. Based
on this assumption the concentration of non-FM Mn ions can
approximately be derived from $M_\mathrm{non-FM}$. For examples,
with $x = 0.2$, the estimated redundant concentration of Mn$^{4+}$
is 0.187 for $M^\prime=\mathrm{Al}$ and the corresponding
concentration of Mn$^{3+}$ is 0.064 for $M^\prime=\mathrm{Ti}$. With
sufficiently high substitution concentrations, contributions to the
the non-FM phase may also arise from the Mn ions of both species
that are isolated by the non-magnetic ones.
\begin{figure}
  % Requires \usepackage{graphicx}
  \vspace{0.1in}
  \includegraphics[width=3.3in]{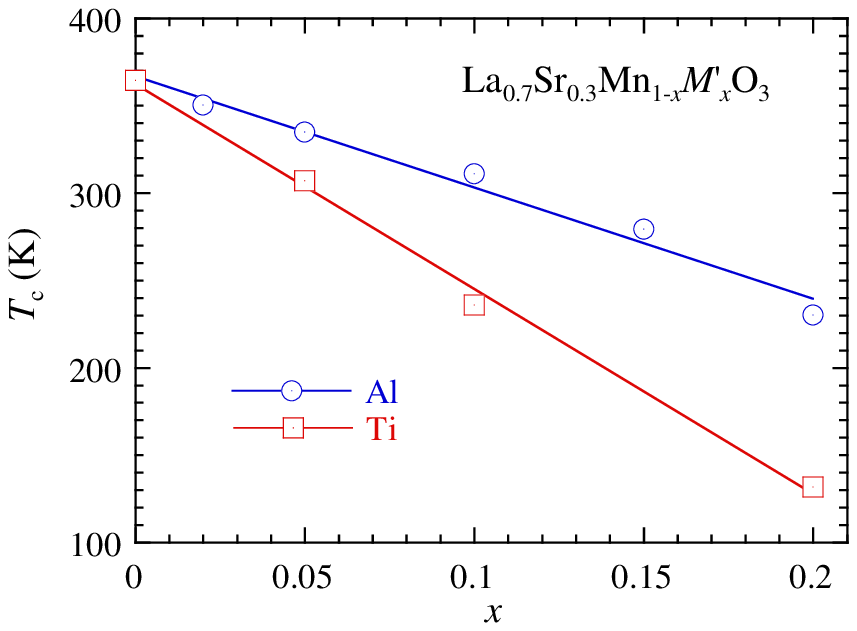}
  \caption{La$_{0.7}$Sr$_{0.3}$Mn$_{1-x}$\textit{M}$^\prime_x$O$_3$:
Variation of $T_\mathrm{c}$ with substitution concentration $x$
($0\leq x\leq0.2$) for $M^\prime=\mathrm{Al}$ ($\odot$) and Ti
($\boxdot$).}
  \label{Figure4}
\end{figure}
\subsection {Effects of selective dilution and the MFT approximation}
Figure \ref{Figure4} shows the dependence of the FM-PM phase
transition temperature $T_\mathrm{c}$ on substitution concentration,
derived from the low-field $M(T)$ data in Fig. \ref{Figure1}. The
$T_\mathrm{c}$ values are determined by the temperatures at which
the $\partial M/\partial T$ curves peak. The parent compound,
$\mathrm{La}_{0.7}\mathrm{Sr}_{0.3}\mathrm{MnO}_3$, has
$T_\mathrm{c} = 364.4$ K, in good agreement with previously reported
data.\cite{Coey} $T_\mathrm{c}$ decreases monotonically as $x$
increases in both cases. Explanations for the decrease of
$T_\mathrm{c}$ with Al and Ti substitution in manganites have been
suggested by several authors. \cite{Sawaki, Qin, Kallel, Kim} One
explanation is simply that the decrease of $T_\mathrm{c}$ is due to
the suppression of long-range FM order of the localized
$t_\mathrm{2g}$ spins by local breakdown of the exchange couplings
where the substitution occurs. \cite{Qin,Sawaki} Hu \textit{et al.}
\cite{Hu} assumed that the substituted Ti$^{4+}$ ions tends to
demolish the DE $\mathrm{Mn}^{3+}-\mathrm{O}^{2-}-\mathrm{Mn}^{4+}$
bonds and lowers the hole carrier concentration, thus suppressing
the DE interaction and lowering $T_\mathrm{c}$. Kallel \textit{et
al.} \cite{Kallel} suggested that the presence of Ti in LSMT$_x$
favors SE interaction and suppresses the DE mechanism.
Significantly, Kim \textit{et al.} \cite{Kim} recently found that
the Ti substitution in LSMT$_x$ increases the
$\mathrm{Mn}-\mathrm{O}-\mathrm{Mn}$ bond length and reduces the
bond angle. Based on structural data, the authors calculated the
variation of $e_\mathrm{g}$-electron bandwidth $W$, finding a
decrease of $W$ with $x$, and related it to the decrease of
$T_\mathrm{c}$. It is worth noting that the ionic size of Mn$^{4+}$
($0.530$ \AA) is smaller than that of Ti$^{4+}$ ($0.605$ \AA) while
Mn$^{3+}$ ($0.645$ \AA) is larger than Al$^{3+}$ ($0.535$
\AA).\cite{Shannon} As a result, the effects of Al and Ti
substitution on the structure, and hence $W$, may not be the same
and even opposite in the two cases. However, in reality,
$T_\mathrm{c}$ has been found always to decrease with the
substitution. A detailed study on the structure of LSMA$_x$ may help
justify the cause.

\begin{figure}
  % Requires \usepackage{graphicx}
  \vspace{0.1in}
  \includegraphics[width=3.3in]{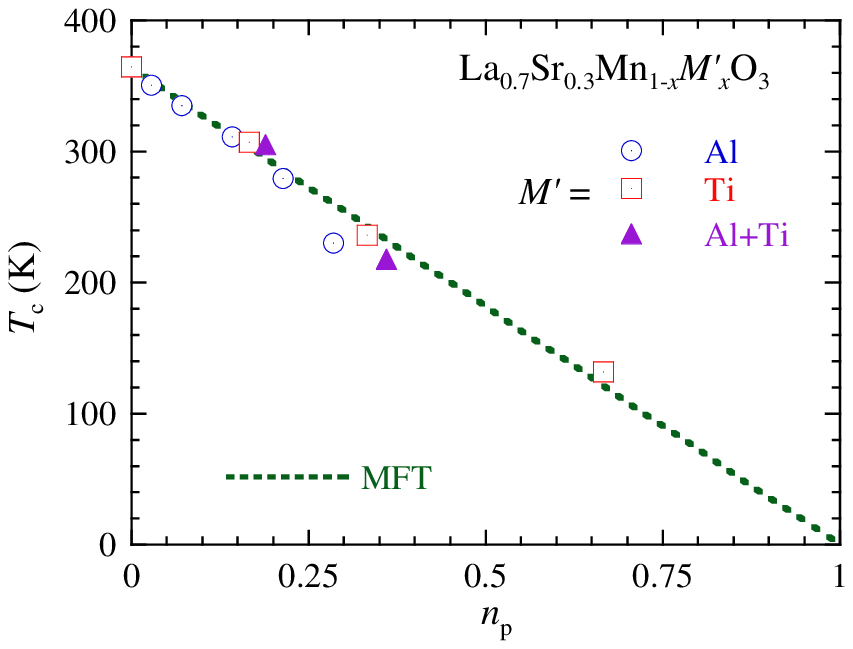}
  \caption{La$_{0.7}$Sr$_{0.3}$Mn$_{1-x}$\textit{M}$^\prime_x$O$_3$:
$T_\mathrm{c}$ is presented as a function of relative concentration
$n_\mathrm{p} = x/0.7$ for $M^\prime=\mathrm{Al}$ ($\odot$) or
$n_\mathrm{p} = x/0.3$ for $M^\prime=\mathrm{Ti}$ ($\boxdot$). The
broken line presents a prediction from molecular-field theory. The
data for $\mathrm{La_{0.7}Sr_{0.3}Mn_{0.9}Al_{0.07}Ti_{0.03}O_{3}}$
($n_\mathrm{p}=0.19$) and
$\mathrm{La_{0.7}Sr_{0.3}Mn_{0.8}Al_{0.14}Ti_{0.06}O_{3}}$
($n_\mathrm{p}=0.36$) ($\blacktriangle$) are added for further
discussions (see text for details).}
  \label{Figure5}
\end{figure}

As is seen in Fig. \ref{Figure4}, the effect of substitution with Ti
on the reduction of $T_\mathrm{c}$ (defined by $\Delta
T_\mathrm{c}=T_\mathrm{c}$($x$)$-T_\mathrm{c}$(0)) is as much as
more than twice of that of Al substitution. However, as mentioned
above, because the substitution is selective, in order to compare
their effects, instead of using $x$ as the common variable, the data
should be presented as functions of relative substitution
concentrations, defined as $n_\mathrm{p} = x/0.7$ when $M^\prime =
\mathrm{Al}$ and $n_\mathrm{p} = x/0.3$ when $M^\prime =
\mathrm{Ti}$. Physically, $n_\mathrm{p}$ is the average
concentration of $M^\prime$ per Mn site of the
selectively-substituted Mn ionic species (Mn$^{3+}$ for $M^\prime =
\mathrm{Al}$, or Mn$^{4+}$ for $M^\prime = \mathrm{Ti}$), or the
probability that site is occupied by an $M^\prime$ ion. Strikingly,
as displayed in Figure \ref{Figure5}, the
$T_\mathrm{c}(n_\mathrm{p})$ data for $M^\prime = \mathrm{Al}$ and
Ti collapse onto one curve which follows very well the linear line
predicted by the molecular-field theory (see below). It is also
surprising that the linear behavior of $T_\mathrm{c}(n_\mathrm{p})$
of LSMT$_x$ sustains in a very wide range of $n_\mathrm{p}$ and has
the tendency to reduce to zero at $n_\mathrm{p}=1$.

For further understanding and analyzes of the results, we use a
mean-field approximation to derive the relation between
$T_\mathrm{c}$ and $n_\mathrm{p}$. According to the Heisenberg
model, the potential energy of exchange interactions of any
particular magnetic ion $i$ with the other ions $j$ is given by $U_i
= -2S_i\cdot\sum_{j\neq i}J_{ij}S_j$ where $S_i$, $S_j$ are the
spins of the $i$th and $j$th ions respectively; and $J_{ij}$ the
exchange integral. $U_i$ can be rewritten as
\begin{equation}
U_i=-2\frac{\mu_i}{g^2}\cdot\sum_{j\neq
i}J_{ij}\mu_j=-2\frac{\mu_i\cdot M}{Ng^2}\sum_{j\neq i}J_{ij}
\label{eqn1}
\end{equation}
where $M$ and $N$ are the magnetic moment and the number of magnetic
ions per unit volume, respectively and  $\mu_i$, $\mu_j$ are the
magnetic moment of the $i$th and $j$th ions. For simplicity, $\mu_j$
is replaced with the average magnetic moment per site
$\langle\mu\rangle=\frac{M}{N}$. Assuming that the $i$th ion
interacts only with its nearest-neighbor ions, but with $n$
different exchange coupling constants $J_{i\alpha}$  each involves
$z_{i\alpha}$ ions, then instead of summing over the ions $j$, the
summation is made over the interactions. We can then recast Eq.
\ref{eqn1} as
\begin{equation}
U_i=-2\frac{\mu_i\cdot
M}{Ng^2}\sum_{\alpha}^{n}z_{i\alpha}J_{i\alpha}. \label{eqn2}
\end{equation}
According to the MFT, $U_i=-\mu_i\cdot B_\mathrm{M}=-\mu_i\cdot
\lambda M$ and $T_\mathrm{c}=\lambda C$, where $B_\mathrm{M}$ is the
molecular field, and $C$ and $\lambda$ are, respectively, the Curie
and Weiss constants, Eq. \ref{eqn2} becomes
\begin{equation}
T_\mathrm{c}=\frac{2S_i\left(S_i+1\right)}{3k_\mathrm{B}}\sum_{\alpha}^{n}z_{i\alpha}J_{i\alpha}.
\label{eqn3}
\end{equation}
For a simple system where there is only one species of magnetic ion
and hence one kind of exchange interaction, Eq. \ref{eqn3} equals
the standard expression
\begin{equation}
T_\mathrm{c}=\frac{2S\left(S+1\right)zJ}{3k_\mathrm{B}}.
\label{eqn4}
\end{equation}
When the system is diluted by a substituting non-magnetic element,
and if the substitution is completely random with respect to its
lattice site, the dilution would result in a proportional dependence
$z(n_\mathrm{p})=z(0)(1-n_\mathrm{p})$ where $0\leq
n_\mathrm{p}\leq1$ and $z(0)$ refers to the undiluted system.
Replacing $z(n_\mathrm{p})$ for $z$ in Eq. \ref{eqn4} yields a
linear dependence of $T_\mathrm{c}$ on $n_\mathrm{p}$ as illustrated
in Figure \ref{Figure5}.

Nevertheless, Eq. \ref{eqn4} is perhaps not relevant to our
manganite systems at first because of the mixed-valence of the Mn
ions and the coexistence of different kinds of interaction including
SE AF interactions
$\mathrm{Mn}^{3+}-\mathrm{O}^{2-}-\mathrm{Mn}^{3+}$,
$\mathrm{Mn}^{4+}-\mathrm{O}^{2-}-\mathrm{Mn}^{4+}$, and the DE FM
interaction $\mathrm{Mn}^{3+}-\mathrm{O}^{2-}-\mathrm{Mn}^{4+}$ with
corresponding exchange constants denoted as $J_{\mathrm{SE1}}$,
$J_{\mathrm{SE2}}$, and $J_{\mathrm{DE}}$, and nearest-neighbor
interacting ions number $z_{\mathrm{SE1}}$, $z_{\mathrm{SE2}}$, and
$z_{\mathrm{DE}}$, respectively. According to Eq. \ref{eqn3}, a
linear behavior is expected to be observed for
$T_\mathrm{c}\left(n_\mathrm{p}\right)$ only if (i) the exchange
coupling constants are unchanged and (ii) the dilution either
affects $z_{i\alpha}$ in a proportional manner such that
$z_{i\alpha}\left(n_\mathrm{p}\right) \propto
\left(1-n_\mathrm{p}\right)z_{i\alpha}(0)$ or does not affect it at
all. Supposing that $J_\mathrm{DE}$, $J_\mathrm{SE1}$,
$J_\mathrm{SE2}$ do not change with substitution, within the
linearity regime of $T_\mathrm{c}(n_\mathrm{p})$, it is presumable
that the Al (or Ti) substitution leaves $z_\mathrm{SE2}$ (or
$z_\mathrm{SE1}$) intact but possibly changes both $z_\mathrm{DE}$
and $z_\mathrm{SE1}$ (or $z_\mathrm{SE2}$) proportionally to
$1-n_\mathrm{p}$. The most remarkable feature of the
$T_\mathrm{c}(n_\mathrm{p})$ variation in Fig. \ref{Figure5} is that
$T_\mathrm{c}$ has a tendency to go to zero when $n_\mathrm{p}=1$.
This would simply mean that the $zJ$ product of DE couplings is
totally dominant over those of the SE ones. Considering the fact
that at low dilution concentrations, since the Mn$^{3+}$ and
Mn$^{4+}$ concentrations are not too different, $z_\mathrm{DE}$,
$z_\mathrm{SE1}$, and $z_\mathrm{SE2}$ are possibly comparable, thus
it is reasonable to suppose that the SE couplings $J_\mathrm{SE1}$,
$J_\mathrm{SE2}$ both are negligibly small in these systems. This
could be one reason behind the fact that
$\mathrm{La}_{0.7}\mathrm{Sr}_{0.3}\mathrm{MnO}_3$ is a unique
manganite in the sense that it has the highest $T_\mathrm{c}$
amongst the perovskite manganites. In a situation when the SE
coupling constants are significant, because either $z_\mathrm{SE1}$
or $z_\mathrm{SE2}$ does not change by selective substitution,
according to Eq. \ref{eqn3}, $T_\mathrm{c}(n_\mathrm{p})$ may still
have a linear dependence but should have a tendency to intercept
$n_\mathrm{p}$-axis at a certain value $n_\mathrm{p}<1$.

Apart from changes of the $z$ values, the substitution of Al or Ti
for Mn could cause some additional effects. Because they are
nonmagnetic, the magnetic coupling will be broken at any site they
occupy, leading to a weakening of long-range FM ordering established
by the dominant DE interactions and a deterioration of the metallic
conductivity. The selectivity of the substitution also drives the
Mn$^{3+}$/Mn$^{4+}$ ratio away from the optimal 7/3 value
contributing to developing the non-FM phase. The differences in
ionic sizes between Mn and Al and Ti could modify the crystal
structure, especially the angle and length of the
$\mathrm{Mn}-\mathrm{O}-\mathrm{Mn}$ couplings.\cite{Garcia} All
these factors contribute to the degradation of the ferromagnetism
and metallicity of the manganite system. Nevertheless, the linear
behavior of $T_\mathrm{c}(n_\mathrm{p})$ observed in Fig.
\ref{Figure5} implies that, within the substitution ranges, those
effects are dominated by the effect of dilution. In addition, there
is a significant difference between the two substitutions. While the
amount of Al substitution reduces the number of hopping electrons by
an equivalent amount, Ti substitution does not affect it at all. The
shortage of hopping electrons would possibly explain the large
number of Mn$^{4+}$ redundancy in the LSMA$_x$ compounds. We believe
that this difference has a link to the reason why the linear
behavior of $T_\mathrm{c}(n_\mathrm{p})$ of LSMA$_x$ occurs in a
much narrower range of dilution ($n_\mathrm{p}\leq0.25$) than that
of LSMT$_x$ where $T_\mathrm{c}(n_\mathrm{p})$ is found linear up to
$n_\mathrm{p} = 0.67$. It is worth noting that
$T_\mathrm{c}(n_\mathrm{p})$ would not follow the linear behavior up
to $n_\mathrm{p} = 1$ as predicted by the MFT because there should
exist a percolation threshold $n_\mathrm{c}$, above which clustering
occurs and the ferromagnetic network collapses into only short-range
ordering and superparamagnetism.

In the case when both Al and Ti are substituted for Mn with
according $n_\mathrm{p}(\mathrm{Al})$ and
$n_\mathrm{p}(\mathrm{Ti})$, supposing that the ferromagnetic double
exchange is totally dominant in the system, the dilution
concentration of the whole system is determined as $n_\mathrm{p} =
1-\left(1-n_\mathrm{p}(\mathrm{Al})\right)\left(1-n_\mathrm{p}(\mathrm{Ti})\right)$.
To check the validity of the MFT analysis in this case, the
$T_\mathrm{c}(n_\mathrm{p})$ values of the
$\mathrm{La_{0.7}Sr_{0.3}Mn_{0.9}Al_{0.07}Ti_{0.03}O_{3}}$
($n_\mathrm{p}=0.19$) and
$\mathrm{La_{0.7}Sr_{0.3}Mn_{0.8}Al_{0.14}Ti_{0.06}O_{3}}$
($n_\mathrm{p}=0.36$) compounds are also added to Figure
\ref{Figure5}. The data fit fairly well the MFT prediction. However,
when the superexchange interactions are taken into account, the
physical scenario for this case will be more complicated and may
need further detailed investigations.
\section{Conclusion}
We have reexamined the magnetic and transport properties of the
La$_{0.7}$Sr$_{0.3}$Mn$_{1-x}M^\prime_x$O$_3$ system, where Mn is
selectively substituted by $M^\prime = \mathrm{Al}$ or Ti; either
Mn$^{3+}$ or Mn$^{4+}$ is selectively substituted for, but
$M^\prime$ is randomly distributed in the Mn network. The analyzes
of the $M(H)$ measurements and saturation magnetization revealed
that the substitution appears to not only merely dilute the magnetic
lattice of Mn ions, but also induce a redundancy of Mn ions. We have
introduced the \emph{selective dilution concentration}
$n_\mathrm{p}$ and discovered that in certain ranges of
$n_\mathrm{p}$, depending on particular $M^\prime$, $T_\mathrm{c}$
scales very well with $n_\mathrm{p}$ in linearity, being in good
agreement with the MFT approximation. The tendency of $T_\mathrm{c}$
to reduce to zero at $n_\mathrm{p}=1$ suggests a dominant role of
the DE mechanism in this system; the SE interaction is effectively
negligible. Remarkably, the linear behavior of
$T_\mathrm{c}(n_\mathrm{p})$ is observed in a very wide range of
$n_\mathrm{p}$ in LSMT$_x$ making this system an excellent candidate
for studies on the effects of dilution where $T_\mathrm{c}$ could be
almost independently tuned without side effects. It is also proposed
that the reduction in the number of hopping $e_\mathrm{g}$ electrons
is one cause for the difference in the effects of dilution between
the LSMT$_x$ and LSMA$_x$ compounds. The MFT analyzes are found to
some extend also valid for the case when Al and Ti are both
substituted for Mn.
\begin{acknowledgments}
This work has been performed partly under the sponsorship of a
collaborative project between the Institute of Materials Science
(VAST, Vietnam) and Uppsala University (Sweden). Two of us would
like to thank the University of Cincinnati and the National Science
Foundation for support. Dr. L. V. Bau would like to acknowledge the
financial support from the Ph.D Training Program of the Ministry of
Education and Training of Vietnam, and is in debt to the
collaboration and training project between the IMS and Hongduc
University.
\end{acknowledgments}

\end{document}